# Hydrogen ($H_2$) Storage in Clathrate Hydrates

*Pratim Kumar Chattaraj,[*] Sateesh Bandaru and Sukanta Mondal*

Department of Chemistry and Center for Theoretical Studies, Indian Institute of Technology,

Kharagpur 721302, India

Author to whom correspondence should be addressed; E-mail: pkc@chem.iitkgp.ernet.in

**Abstract**

Structure, stability and reactivity of clathrate hydrates with or without hydrogen encapsulation are studied using standard density functional calculations. Conceptual density functional theory based reactivity descriptors and the associated electronic structure principles are used to explain the hydrogen storage properties of clathrate hydrates. Different thermodynamic quantities associated with $H_2$-trapping are also computed. The stability of the $H_2$-clathrate hydrate complexes increases upon the subsequent addition of hydrogen molecules to the clathrate hydrates. The efficacy of trapping of hydrogen molecules inside the cages of clathrate hydrates depends upon the cavity sizes and shapes of the clathrate hydrates. Computational studies reveal that **$5^{12}$** and **$5^{12}6^2$** structures are able to accommodate up to two $H_2$ molecules whereas **$5^{12}6^8$** can accommodate up to six hydrogen molecules.

**Introduction**

Hydrogen is an extremely environment friendly fuel and it is viewed as a promising clean fuel of the future. When it burns it releases only water vapor into the environment, unlike fossil fuels such as oil, natural gas and coal. Those which contain carbon, produce $CO_2$, CO, including a higher ratio of carbon emissions, nitrogen oxides ($NO_x$), and sulfur dioxide ($SO_2$). Coal and fuel oil also release ash particles into the environment, substances that do not burn but instead are carried into the atmosphere and contribute to pollution and global warming and are available in limited supply. Hydrogen is clean, highly abundant, and non-toxic, renewable fuel and also packs energy per unit mass like any other expensive fuels. The major problem with this fuel is its storage, because it needs to be stored like other compressed gases. Hydrogen is usually stored in three different ways, namely compression, liquefaction, and storage in a solid material.[1, 2]

A few of the research groups have reported novel materials for storage of hydrogen at ambient conditions.[3-5] Materials, such as aluminum nitride (AlN) nanostructures,[6] transition-metal doped boron nitride (BN) systems,[7] alkali-metal doped benzenoid[8] and fullerene clusters,[9] light metal and transition-metal coated boron buckyballs, $B_{80}$,[10] and magnesium clusters[11] ($Mg_n$) have been confirmed both experimentally and theoretically to serve as potential hydrogen-storage materials. Further, based on a theoretical study invoking the metastability of hydrogen-stacked $Mg_n$ clusters,[12] our group has very recently demonstrated that a host of small to medium sized metal cluster moieties involving $Li_3^+$, $Na_3^+$, $Mg_n$ and $Ca_n$ (n = 8-10) cages have got a fair capability of trapping hydrogen in both atomic and molecular forms.[13]

In recent years, clathrate hydrates have generated much interest in the study of the hydrogen storage materials. Clathrate hydrates are inclusion compounds which incorporate guest molecules inside the polyhedral cages of the host framework made up of hydrogen-bonded water molecules.[14] Clathrate hydrates generally form two cubic structures: type I, type II[15] and rarely, a third hexagonal cubic structure as type H[16] consisting of cages of water molecules. Each structure has different crystallographic properties and contains cavities of different shapes and sizes. In case of type I, a clathrate consists of 46 water molecules that form two pentagonal dodecahedron ($5^{12}$) and six hexagonal truncated trapezohedron ($5^{12}6^2$) cages in a unit cell. In case of type II, it consists of 136 water molecules that form sixteen $5^{12}$ and eight $5^{12}6^4$ cages in a unit cell. The type H clathrate consists of 36 water molecules that form three $5^{12}$, two $4^35^66^3$, and one $5^{12}6^8$ cages in a unit cell. These clathrate hydrates act as host complexes and in which the host molecules are water and the guest molecules are gases and without the support of the trapped gas molecules, the structure of hydrate structures would collapse into liquid water.

Recently, clathrate hydrates have been synthesized with hydrogen molecules as guests at high pressures (~200 Mpa) and low temperatures (~250 K).[16] Hydrates have been studied for a variety of applications including: energy production,[17-18] deposition of methane hydrates,[19] water desalination,[20] $CO_2$ sequestration,[21] and gas separation.[22] Based on first-principle quantum chemistry calculations, Tse et al. found that the stability of the hydrogen clathrate is mainly due to the dispersive interactions between the molecules of $H_2$ and the water molecules of hydrates.[23] In the same study they explained that hydrogen molecules undergo essentially free rotations inside the clathrate cages. Mao et al. performed[14] studies based on high-pressure Raman, infrared, x-ray, and neutron studies which show that $H_2$ and $H_2O$ mixtures crystallize into the sII clathrate structure with an approximate $H_2/H_2O$ molar ratio of 1:2. The clathrate cages are occupied by multiple members, with a cluster of two $H_2$ molecules ($5^{12}$) in the small cage and four in the large cage ($5^{12}6^4$). The density functional calculations reported so far, suggest that small cages can accept only two hydrogen molecules. This work is probably the first comprehensive computational study of the storage of hydrogen in clathrate hydrates.

In the present study computations are carried out to explain the binding ability of hydrogen towards the various clathrate hydrate structures. The objective of this study is to find out how many hydrogen molecules are possible to be stored in each clathrate hydrate cavity, upon subsequent addition of hydrogen. The binding ability is explained based on interaction energies and other thermodynamic properties. Stability and reactivity of the clathrate hydrates with and without $H_2$ complexes can be explained based on conceptual density functional theory (DFT) based reactivity descriptors.

**Theoretical Background**

The stability of clathrate hydrates-$nH_2$ systems may be justified from their chemical hardness ($\eta$) and electrophilicity ($\omega$) values. This has been further validated by the establishment of some associated molecular electronic structure principles like the principle of maximum hardness[24-26] (PMH) together with the minimum polarizability principle[27-28] (MPP) and minimum electrophilicity principle[29-30] (MEP). These electronic structure principles serve as major determinants towards assessing the stability and reactivity trends of chemical systems. For an N-electron system, the electronegativity[31-33] ($\chi$) and hardness[34-36] ($\eta$) can be defined as follows:

$$\chi = -\mu = -\left(\frac{\partial E}{\partial N}\right)_{v(\vec{r})} \quad (1)$$

and

$$\eta = \left(\frac{\partial^2 E}{\partial N^2}\right)_{v(\vec{r})} \quad (2)$$

Here $E$ is the total energy of the *N*-electron system and $\mu$ and $v(\vec{r})$ are its chemical potential and external potential respectively. The electrophilicity[37-39] ($\omega$) is defined as:

$$\omega = \frac{\mu^2}{2\eta} = \frac{\chi^2}{2\eta} \quad (3)$$

A finite difference approximation to Eqs.1 and 2 can be expressed as:

$$\chi = \frac{I+A}{2} \qquad (4)$$

and

$$\eta = I - A \qquad (5)$$

where $I$ and $A$ represent the ionization potential and electron affinity of the system respectively.

**Computational Details**

All the clathrate hydrate structures, $5^{12}$, $5^{12}6^2$, $5^{12}6^4$ and $5^{12}6^8$ with and without $H_2$ complexes considered in this study are optimized using the gradient corrected hybrid density functional B3LYP in conjunction with 6-31G(d) basis set. Frequency calculations reveal that $5^{12}$, $5^{12}6^2$ and $5^{12}6^8$ clathrate hydrate structures are minima on the potential energy surface without any imposition of the symmetry constraints and for the structure $5^{12}6^4$ we have not obtained the minimum. For this cage structure we first freeze the "O" atom coordinates and do the optimization at B3LYP/6-31G(d) level of theory. The electrophilicity ($\omega$) and hardness ($\eta$) are computed using the eqs.3 and 5 respectively. Single point calculations are also done using 6-31+G(d) basis set on the B3LYP/6-31G(d) optimized geometries. We also calculate the changes in enthalpy and Gibbs' free energy associated with encapsulation. Interaction energies ($\Delta E_r$) are calculated using the following formula,

$$\Delta E_r = [E_{C-H_2} - (E_{H_2} + E_C)] \qquad \ldots\ldots\ldots(6)$$

where the subscripts C-$H_2$, H and $C$ represent the total energies of the complex, $H_2$ and cavity structure respectively.

All the calculations are done using the Gaussian 03 suite of program.[40]

In case of **$nH_2@5^{12}$**, we carry out freezing optimization on $5^{12}$ cage structure as well as $H_2$ encapsulated cage structures. In this case we first freeze the "O" atom coordinates and perform the geometry optimization at B3LYP/6-31G(d) level of theory. Single point energy calculations at the MP2 level are performed using 3-21G basis set augmented by 2p polarization functions on hydrogen atoms. All these single point calculations are performed by using GAMESS program[41] on DFT optimized frozen geometries. For **nH$_2$@5$^{12}$** normal systems (without constraints) we perform MP2/3-21G single point energy calculations on DFT optimized geometries.

**Results and Discussion**

Among all these major building blocks (Type I, Type II and Type H) we consider four different clathrate hydrates as host molecules namely **5$^{12}$, 5$^{12}$6$^{2}$, 5$^{12}$6$^{4}$** and **5$^{12}$6$^{8}$** for the study of the hydrogen (H$_2$) interaction ability of clathrate hydrates and the calculation of quantum chemical descriptors. The notation **5$^{12}$** stands for a cavity with 12 pentagonal faces. The notation **5$^{12}$6$^{k}$** (k = 2, 4, 8) stands for a cavity with 12 pentagonal and k hexagonal faces.

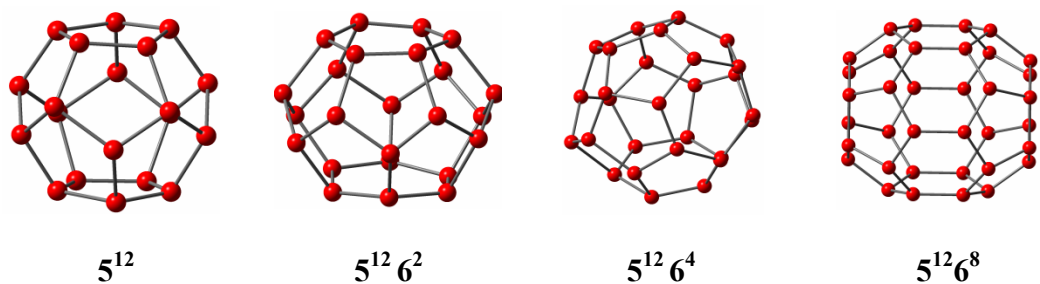

**5$^{12}$**   **5$^{12}$ 6$^{2}$**   **5$^{12}$ 6$^{4}$**   **5$^{12}$6$^{8}$**

**Scheme 1**. Structures considered for the study of hydrogen encapsulation (Hydrogens are omitted for the sake of clarity)

For the small clathrate cage **5$^{12}$**, we consider complexes encapsulated with one to six hydrogen molecules. Among the six H$_2$-cluster complexes, we are able to find the stationary points up to five hydrogen molecules. Addition of six hydrogen molecules to **5$^{12}$** cavity results in the structural deformation and the minimum energy structure for the corresponding complex is not achieved at the B3LYP/6-31G(d) level of theory. From the frequency calculations we find that all five H$_2$-cluster complexes are minima on the potential energy surface. We calculate the interaction energies using Eq-6. Addition of hydrogen molecules (H$_2$) to clathrate hydrate structures (C) are treated as a series of chemical reactions shown below.

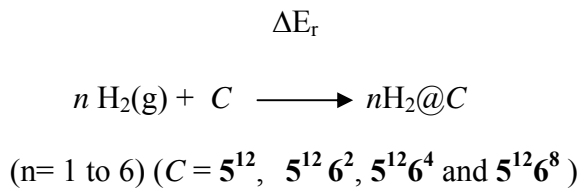

$$n\,H_2(g) + C \xrightarrow{\Delta E_r} nH_2@C$$

(n= 1 to 6) (C = $5^{12}$, $5^{12}6^2$, $5^{12}6^4$ and $5^{12}6^8$)

**TABLE 1:** Energy (E, au), Interaction energies ($\Delta E_r$, kcal/mol), Free energy change ($\Delta G$, kcal/mol) and Enthalpy change ($\Delta H$, kcal/mol) of $nH_2$ (1-6) Encapsulation to $5^{12}$ at B3LYP/6-31G(d) and Interaction energies (kcal/mol) at the MP2/6-31G(d)(*$\Delta E_r$) and MP2/6-31+G(d) Level of Theory.

| Structure | Energy | $\Delta E_r$ | $\Delta G$ | $\Delta H$ | *$\Delta E_r$ | **$\Delta E_r$ | ¶$\Delta E_r$ | $\$\Delta E_r$ |
|---|---|---|---|---|---|---|---|---|
| $1H_2@5^{12}$ | -1529.82768 | -1.08 | 5.45 | 0.17 | -1.68 | -1.60 | -2.64 (-1.84)§ | -3.89 |
| $2H_2@5^{12}$ | -1530.99996 | 0.91 | 16.80 | 4.26 | -0.45 | 0.39 | -9.16 (-9.00)§ | -6.41 |
| $3H_2@5^{12}$ | -1532.16853 | 5.28 | 32.27 | 10.94 | 3.27 | 3.89 | -10.84 (-6.09)§ | -4.97 |
| $4H_2@5^{12}$ | -1533.33319 | 12.07 | 49.95 | 20.29 | 9.20 | 10.32 | -7.99 | -0.72 |
| $5H_2@5^{12}$ | -1534.49045 | 23.45 | 71.61 | 34.63 | 19.50 | 20.82 |  | 11.7 |
| $6H_2@5^{12}$ | -a- | -a- | -a- | -a- | -a- |  | -a- |  |

-a- = The minima are not obtained   **MP2/6-31+G(d)//B3LYP/6-31G(d)

¶ = MP2//DFT interaction energies on freezing geometries

$ = MP2//DFT interaction energies on normal geometries

§ = MP2/DFT literature reported interaction energies[23]

From the Table 1, the encapsulation of first hydrogen molecule to cavity is a favorable process. The encapsulation of successive hydrogen molecules to the cavity is not favorable at the B3LYP/6-31G(d) level of theory. Interaction energies, free energies and enthalpy changes are shown in Table 1. In case of **$nH_2@5^{12}$**, we perform geometry optimization as well as frequency calculations at second order MP2 level of theory using 6-31G(d) basis set and MP2/6-31+G(d) single point energy calculations on B3LYP/6-31G(d) optimized geometries to see dependence of the interaction energies on the method used. From the Table 1 we see that the interaction energies at B3LYP and MP2 levels do not vary much, and both levels give similar tendency for one and two hydrogen molecules, for encapsulation of three, four and five hydrogen molecules to $5^{12}$ cavity, the interaction energy variation in both the methods are approximately 2 kcal/mol.

In case of $6H_2@5^{12}$ system the minimum is not obtained, encapsulation of six hydrogen molecules to $5^{12}$ cavity leads to structural deformation, results in collapse of the structure and moving away of the hydrogen molecule from the cavity. Because of its smaller size, the $5^{12}$ can accommodate up to five hydrogen molecules, and the calculated B3LYP formation enthalpy for the $4H_2@5^{12}$ and $5H_2@5^{12}$ are 20.29 and 34.63 kcal/mol respectively.

*Ab initio* calculations performed by Patchkovskii et al revealed the stable occupancy of two $H_2$ molecules in the small cavity ($5^{12}$) of the clathrate-hydrate species.[23] Later DFT studies by Sluiter et al. on pure $H_2$-hydrates showed that the small cage-cavities can accommodate up to two $H_2$ molecules.[42] In view of the above argument regarding the encapsulation of $H_2$ in the small cage-cavity, we perform the freezing optimization on $5^{12}$ cage, and $H_2$ encapsulated cage structures at the B3LYP/6-31G(d) level of theory. Single-point calculations are performed at the MP2 level by using 3-21G basis set augmented by 2p polarization functions on hydrogen atoms. The interaction energies of all the $nH2@5^{12}$ (n=1-4) complexes both with constraints ($^{¶}\Delta E_r$) and without constraints ($^{\$}\Delta E_r$) are presented in Table 1. Table 1 reveals that the encapsulations of up to four hydrogen molecules are favorable in both the cases. Our calculated interaction energy values corroborate nicely with analogous reports[23] for encapsulation of up to two hydrogen molecules. The incorporation of up to three (-10.84 kcal/mol) as well as four (-7.99 kcal/mol) hydrogen molecules onto the clathrate-cavity are also energetically favorable processes.

For the $nH_2@5^{12}$ clusters, encapsulation of first hydrogen molecule on to the cage cavity shows that the $H_2$ molecule prefers to stay at approximately 0.8Å away from the center of the cage. In case of encapsulation of two $H_2$ molecules at a time onto the bare $5^{12}$ cage-cavity, both the hydrogen molecules prefer to stay perpendicular to each other at a distance of 2.580Å. In the same manner the encapsulation of three hydrogen molecules leads to almost a triangular alignment amongst the H$_2$-moieties separated at distances of 2.581Å, 2.555Å and 2.537Å respectively. The encapsulation of four H$_2$ molecules prefer the tetrahedral shape and of five hydrogen molecules approximately adopt a trigonal bi-pyramidal arrangement.

The associated molecular electronic structure principles play a major role in determining the changes in stability of the clathrate hydrate clusters from their free, unbound state to the corresponding H$_2$ encapsulated form. The total energy (E, a.u.) and the important global reactivity descriptors like electronegativity ($\chi$, eV), hardness ($\eta$, eV) and electrophilicity ($\omega$, eV) ($\omega$) for the H$_2$ encapsulation to **5$^{12}$** hydrate structures are displayed in Table 2.

**TABLE 2:** Electronegativity ($\chi$, eV), Hardness ($\eta$, eV) and Electrophilicity ($\omega$, eV) of **5$^{12}$** and H$_2$-Encapsulated **5$^{12}$** cage structures at B3LYP/6-31G(d) and MP2/6-31+G(d) Level of Theory.

|  | B3LYP/6-31G(d) | | | MP2/6-31G+G(d) | | |
| --- | --- | --- | --- | --- | --- | --- |
| Structure | $\chi$ | $\eta$ | $\omega$ | $\chi$ | $\eta$ | $\omega$ |
| **512** | 2.774 | 6.854 | 0.561 | 5.105 | 15.690 | 0.831 |
| **1H2@512** | 2.762 | 6.898 | 0.553 | 5.030 | 15.839 | 0.799 |
| **2H2@512** | 2.725 | 6.943 | 0.535 | 5.013 | 15.928 | 0.789 |
| **3H2@512** | 2.703 | 7.051 | 0.518 | 4.976 | 15.958 | 0.776 |
| **4H2@512** | 2.678 | 7.066 | 0.507 | 4.993 | 15.995 | 0.779 |
| **5H2@512** | 2.665 | 7.085 | 0.501 | 4.976 | 15.959 | 0.776 |

From the Table 2, it transpires that the stability of **nH2@ 5$^{12}$** complexes increases upon incorporation of hydrogen molecules into the cage-cavity. The hardness of **5$^{12}$** hydrate is 6.854 eV. Upon encapsulation of the hydrogen to cavity the hardness increases, indicating a potential increase in the rigidity of the H$_2$-bound complexes thereby increasing their stability. Figure 1 reveals that the total energy, electronegativity and electrophilicity of the **nH2@5$^{12}$** complexes decrease upon encapsulation of hydrogen to the **5$^{12}$** cavity. An analogous trend for all the global reactivity descriptors for the H$_2$-bound $5^{12}$ hydrate systems are observed at the MP2/6-31+G(d) level as depicted in Table 2. From the global reactivity descriptors, it is evident that upon addition of hydrogen molecules to $5^{12}$ cavity, reactivity of $5^{12}$ structure decreases with an increase in the stability.

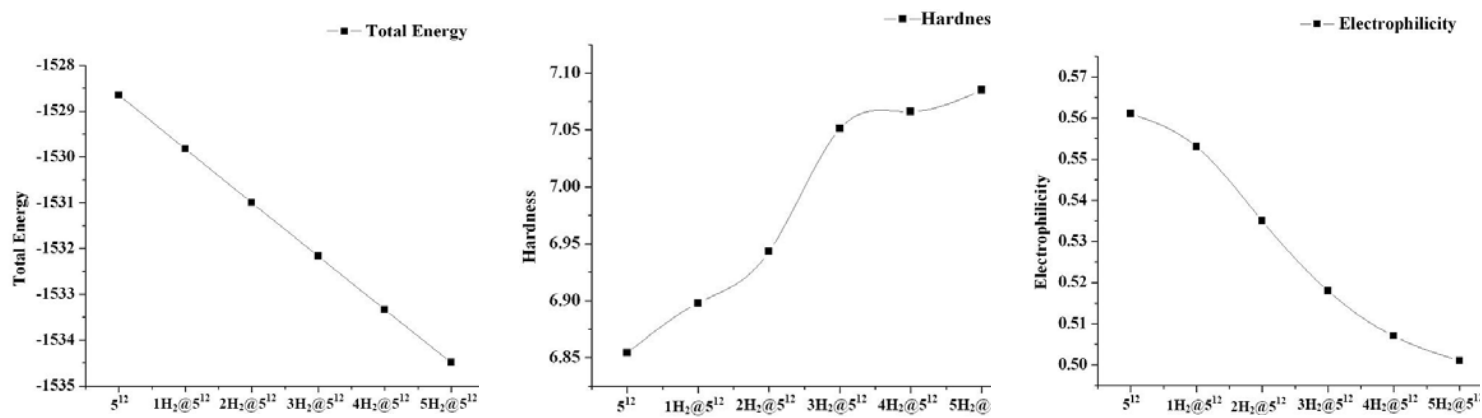

**Figure 1.** Variation of total energy, hardness (η) and electrophilicity (ω) for encapsulation of $nH_2$ to $5^{12}$ clathrate cavity

In case of $5^{12}$ clathrate hydrates cage structure, we perform a total energy scan. The $H_2$ molecule was allowed to migrate stepwise up to 7Å from the center of the cavity and the energies are calculated correspondingly at 0.5Å incremental steps at the B3LYP/6-311+G(d, p) level of theory. A graphical illustration between the distance and the total energy shows that at a distance of 3Å from the cage center the energy achieves the maximum value. As the distance from center of the cavity increases another lower energy structure is found which is called as reactant pair complex.

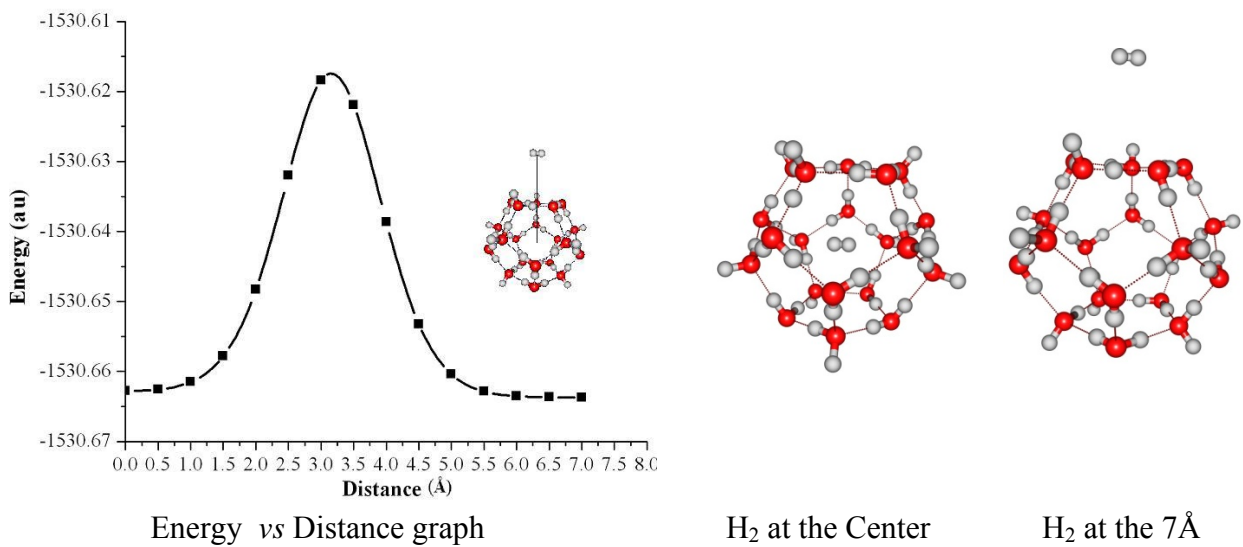

Energy *vs* Distance graph      H₂ at the Center      H₂ at the 7Å

**Figure 2.** Total energy of a single H$_2$ molecule along the line connecting the center of a **5$^{12}$** cage through the center of one pentagonal face to a point which is 7Å apart from the center.

We take both the structures (Figure 2) and performed the geometry optimization and subsequent frequency calculations at the B3LYP/6-31G(d) level of theory. The relative energy difference between the two structures is 0.695 kcal/mol. Hydrogen placed at centroid of the cavity is found to correspond with the real minimum on the potential energy surface whereas the other structure bears two imaginary frequencies. It is therefore fairly clear that as the global minimum for trapping of H$_2$ by the **5$^{12}$** cage lies within the cavity as evident from the energy scan (rigid). It may be concluded that the hydrogen molecule would like to be bound inside the cavity instead of being trapped outside the cage.

In case of the **5$^{12}$6$^2$** clathrate cage, we study the complexes with encapsulation of up to six hydrogen molecules. Among these complexes we are able to find the stationary points of incorporation of up to five hydrogen molecules having a subsequent zero imaginary frequency. But **5$^{12}$6$^2$** clathrate hydrate cage shows profound deformation upon trapping the sixth H$_2$ molecule onto the cage cavity. All calculations are performed at the B3LYP/6-31G(d) level of theory. In case of the encapsulation of up to five hydrogen molecules onto the **5$^{12}$6$^2$** cavity we further notice a little distortion in the structure (**5$^{12}$6$^2$**) which is particularly more pronounced during incorporation of up to three H$_2$ molecules and gradually decreases for the encapsulation of the fourth or fifth hydrogens.

**TABLE 3:** Energy (E, au), Interaction energies ($\Delta E_r$, kcal/mol), Free energy change ($\Delta G$, kcal/mol) and Enthalpy change ($\Delta H$, kcal/mol) of $nH_2$ (1-6) Encapsulation to $5^{12}6^2$ at B3LYP/6-31G(d) and Interaction energies (kcal/mol) at the MP2/6-31+G(d) Level of Theory.

| Structure | Energy (E, au) | $\Delta E_r$ | $\Delta G$ | $\Delta H$ | *$\Delta E_r$ |
|---|---|---|---|---|---|
| **1H$_2$@5$^{12}$6$^2$** | -1835.55058 | -3.05 | 6.19 | -1.49 | -3.03 |
| **2H$_2$@5$^{12}$6$^2$** | -1836.72586 | -2.92 | 13.84 | 0.05 | -3.07 |
| **3H$_2$@5$^{12}$6$^2$** | -1837.89459 | 1.32 | 26.04 | 5.72 | -1.21 |
| **4H$_2$@5$^{12}$6$^2$** | -1839.06545 | 4.22 | 39.39 | 10.74 | 2.13 |
| **5H$_2$@5$^{12}$6$^2$** | -1840.23208 | 9.78 | 53.71 | 18.37 | 6.56 |
| **6H$_2$@5$^{12}$6$^2$** | -a- | -a- | -a- | -a- | -a- |

-a- The minima are not obtained *MP2/6-31+G(d)//B3LYP/6-31G(d)

From the Table 3 it is quite evident that while the encapsulation of one and two hydrogen molecules onto $5^{12}6^2$ cavity is an energetically favorable process owing to a favorable interaction energy values, further encapsulation of three, four or five hydrogen molecules to $5^{12}6^2$ cavity, however, shows a slightly conflicting trends in the interaction energies. For all the **nH$_2$@5$^{12}$6$^2$** complexes, the total energy, interaction energies, free energies and enthalpy changes are given in Table 3. Analogous studies at the MP2/6-31+G(d) level show an energetically favorable encapsulation of up to three hydrogen molecules onto the $5^{12}6^2$ cavity.

**TABLE 4:** Electronegativity ($\chi$, eV), Hardness ($\eta$, eV) and Electrophilicity ($\omega$, eV) of $5^{12}6^2$ and $H_2$-Encapsulated $5^{12}6^2$ cage structures at B3LYP/6-31G(d) and MP2/6-31+G(d) Level of Theory.

| Structure | B3LYP/6-31G(d) | | | MP2/6-31+G(d) | | |
|---|---|---|---|---|---|---|
| | $\chi$ | $\eta$ | $\omega$ | $\chi$ | $\eta$ | $\omega$ |
| $5^{12}6^2$ | 3.495 | 7.545 | 0.809 | 5.272 | 15.132 | 0.918 |
| $1H_2@5^{12}6^2$ | 3.437 | 7.753 | 0.762 | 5.186 | 15.569 | 0.864 |
| $2H_2@5^{12}6^2$ | 3.464 | 7.838 | 0.765 | 5.179 | 15.630 | 0.858 |
| $3H_2@5^{12}6^2$ | 3.336 | 7.824 | 0.711 | 5.081 | 15.609 | 0.827 |
| $4H_2@5^{12}6^2$ | 3.343 | 7.890 | 0.708 | 5.096 | 15.656 | 0.829 |
| $5H_2@5^{12}6^2$ | 3.325 | 7.890 | 0.701 | 5.085 | 15.655 | 0.826 |

For the $5^{12}$ and $5^{12}6^2$ hydrogen bound clathrate hydrates the hardness ($\eta$) values are found to correlate nicely with the corresponding electrophilicity ($\omega$) values as evident from Tables 2 and 4 respectively. For the $5^{12}$ and $5^{12}6^2$ hydrogen bonded complexes on an average the hardness ($\eta$) increases with a gradual decrease in the consecutive electrophilicity ($\omega$) values, thereby corroborating the associated principles of maximum hardness and minimum electrophilicity.

The hardness ($\eta$) and electrophilicity ($\omega$) of $5^{12}6^2$ are 7.545 eV and 0.809 eV respectively. For encapsulation of the hydrogen molecule on to the $5^{12}6^2$ cavity, the hardness increases with a subsequent decrease in the electrophilicity values except for the $3H_2@5^{12}6^2$ system as shown in Figure 3. We also present in Figure 3 the total energy plot for encapsulation of the hydrogen molecule onto the $5^{12}6^2$ cavity.

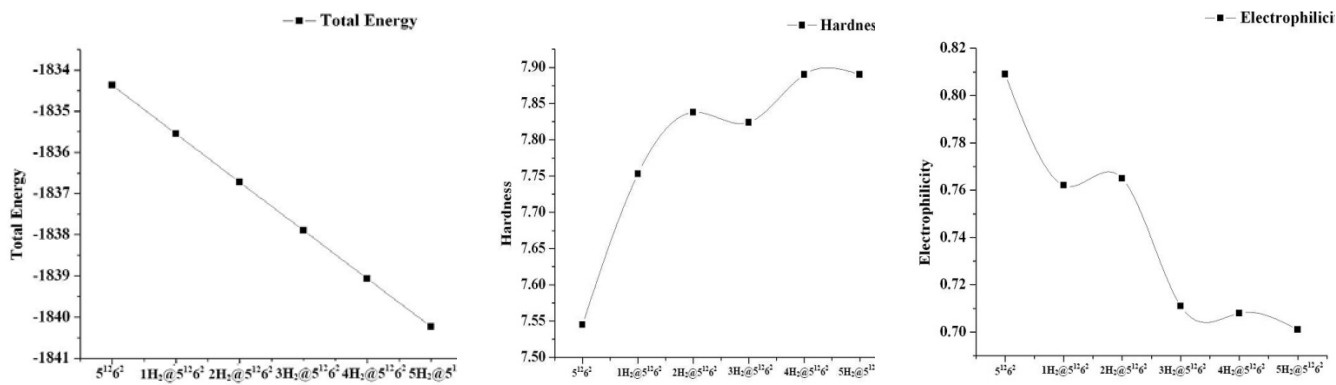

**Figure 3.** Variation of total energy, hardness (η) and electrophilicity (ω) for encapsulation of nH$_2$ to **5$^{12}$6$^2$** clathrate cavity

In case of the first order encapsulation of hydrogen molecule onto the **5$^{12}$6$^2$** cage, the hydrogen molecule prefers to stay near the center of the cage cavity, and for the addition of two hydrogen molecules they prefer to stay at a distance of 3.227Å and the hydrogen molecules are slightly perpendicular to each other. In case of the encapsulation of three hydrogen molecules to cavity hydrogen molecules prefer to stay in a trigonal fashion at distances of 3.214Å, 2.887Å and 2.758Å respectively. In case of addition of four hydrogen molecules to **5$^{12}$6$^2$** cavity, hydrogen molecules prefer to stay in distorted tetrahedral shape, whereas in case of five hydrogen molecules, the H$_2$ arrangement inside the cavity assumes a distorted trigonal bi-pyramidal shape.

In case of **5$^{12}$6$^4$**, the minimum energy structures are not found and all our efforts ultimately lead to a structural distortion. For this cage structure we first freeze the "O" atom coordinates and do the geometry optimization at the B3LYP/6-31G(d) level of theory. Interestingly, encapsulation of a single hydrogen molecule into the cavity of the **5$^{12}$6$^4$** cage stabilizes the structure as computed at B3LYP/6-31G(d) level of theory. From the frequency calculations, we found that the **1H$_2$@5$^{12}$6$^4$** complex is not a minimum on the PES. The interaction energy of **1H$_2$@5$^{12}$6$^4$** complex is -0.38 kcal/mol.

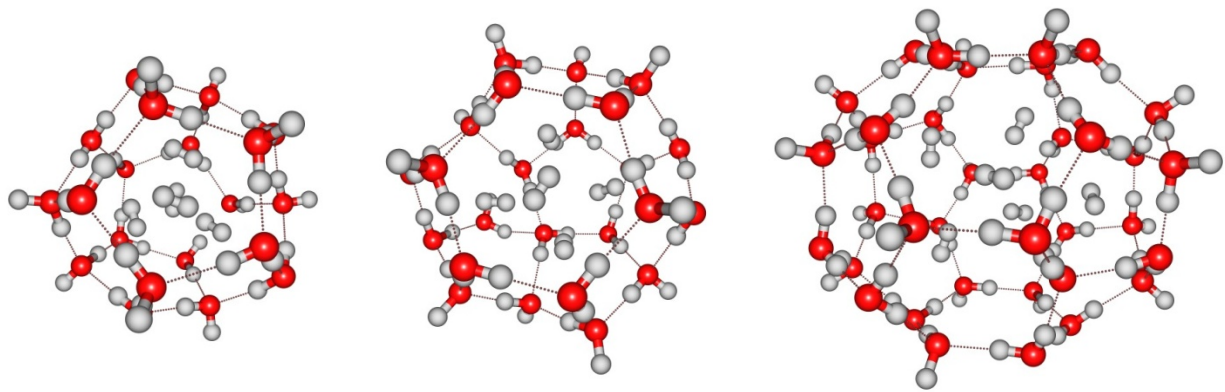

                     $5H_2@5^{12}$                      $5H_2@5^{12}6^2$                      $6H_2@5^{12}6^8$

**Figure 4.** Maximum possibility of hydrogen encapsulation to $5^{12}$, $5^{12}6^2$ and $5^{12}6^8$ optimized structures at B3LYP/6-31G(d) level of theory.

Further attempts to incorporate two, three and four hydrogen molecules onto the $5^{12}6^4$ cage lead to complete structural deformation.

**TABLE 5:** Energies (E, au), Interaction energies ($\Delta E_r$, kcal/mol), Free energy change ($\Delta G$, kcal/mol) and Enthalpy change ($\Delta H$, kcal/mol) of $nH_2$ (1-6) Encapsulation to $5^{12}6^8$ at B3LYP/6-31G(d) and Interaction energies (kcal/mol) at the MP2/6-31+G(d) Level of Theory.

| Structure | Energy (E, au) | $\Delta E_r$ | $\Delta G$ | $\Delta H$ | *$\Delta E_r$ |
|---|---|---|---|---|---|
| $1H_2@5^{12}6^8$ | -2752.73664 | -1.24 | 5.64 | 0.05 | -0.989 |
| $2H_2@5^{12}6^8$ | -2753.91301 | -1.80 | 11.89 | 0.66 | -1.675 |
| $3H_2@5^{12}6^8$ | -2755.09168 | -3.80 | 18.27 | 0.34 | -2.829 |
| $4H_2@5^{12}6^8$ | -2756.26764 | -4.10 | 25.33 | 1.27 | -3.427 |
| $5H_2@5^{12}6^8$ | -2757.44278 | -3.89 | 32.69 | 2.84 | -3.426 |
| $6H_2@5^{12}6^8$ | -2758.61878 | -4.21 | 39.59 | 3.97 | -4.010 |

*MP2/6-31+G(d)//B3LYP/6-31G(d)

In case of $5^{12}6^8$, we consider complexes encapsulated with one to six hydrogen molecules. The nH$_2$ encapsulated $5^{12}6^8$ cluster structures are minima on the potential energy surface. Interestingly from the interaction energies, encapsulation of all hydrogen molecules (n=1-6) to $5^{12}6^8$ cavity is energetically favorable and corresponding interaction energies are given in Table 5. From Table 5, we see that the interaction energies decrease upon encapsulation of hydrogen molecules onto the $5^{12}6^8$ cavity. This is due to the larger size of $5^{12}6^8$ cavity so that it can accommodate up to six hydrogen molecules. Enthalpy change of hydrogen encapsulation to $5^{12}6^8$ structure predicts a slightly endothermic process. Since ΔG value is positive in all cases the complexes are at most kinetically stable. Total energies, interaction energies, free energies and enthalpy changes of all **nH$_2$@5$^{12}$6$^8$** complexes are presented in Table 5.

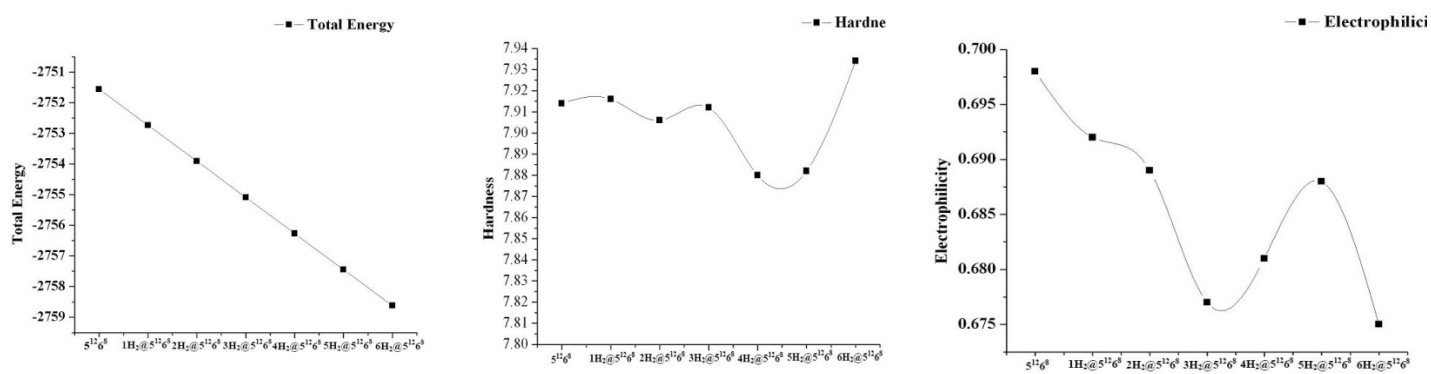

**Figure 5.** Variation of Total energy, hardness (η) and electrophilicity (ω) for encapsulation of nH$_2$ to $5^{12}6^8$ clathrate cavity

The maximum possibility of hydrogen molecules stored in $5^{12}$, $5^{12}6^2$ and $5^{12}6^8$ is shown in Figure 4. Based on these results we conclude that the $5^{12}$ and $5^{12}6^2$ structures, may accommodate up to two hydrogen molecules and $5^{12}6^8$ may store up to six hydrogen molecules.

**TABLE 6:** Electronegativity (χ, eV), Hardness (η, eV) and Electrophilicity (ω, eV) of $5^{12}6^8$ and H$_2$-Encapsulated $5^{12}6^8$ cage structures at B3LYP/6-31G(d) and MP2/6-31+G(d) Level of Theory.

| Structure | B3LYP/6-31G(d) | | | MP2/6-31G+G(d) | | |
|---|---|---|---|---|---|---|
| | χ | η | ω | χ | η | ω |
| $5^{12}6^8$ | 3.324 | 7.914 | 0.698 | 5.177 | 15.357 | 0.872 |
| 1H$_2$@$5^{12}6^8$ | 3.309 | 7.916 | 0.692 | 5.152 | 15.388 | 0.862 |
| 2H$_2$@$5^{12}6^8$ | 3.301 | 7.906 | 0.689 | 5.124 | 15.428 | 0.851 |
| 3H$_2$@$5^{12}6^8$ | 3.274 | 7.912 | 0.677 | 5.058 | 15.515 | 0.825 |
| 4H$_2$@$5^{12}6^8$ | 3.276 | 7.880 | 0.681 | 5.030 | 15.553 | 0.814 |
| 5H$_2$@$5^{12}6^8$ | 3.292 | 7.882 | 0.688 | 5.015 | 15.576 | 0.807 |
| 6H$_2$@$5^{12}6^8$ | 3.273 | 7.934 | 0.675 | 4.997 | 15.614 | 0.800 |

From Table 6, we observe that the stability of H$_2$ encapsulated $5^{12}6^8$ cluster complexes increases upon addition of hydrogen molecules. The hardness of $5^{12}6^8$ is 7.914 eV. Upon encapsulation of the hydrogen to cage-cavity, the hardness increases. However, there is a little deviation in η for the addition of three, four and five hydrogen molecules with a slightly dipping trend. Finally for the addition of six hydrogen molecules to $5^{12}6^8$ cage, the hardness increases again. The total energy and electrophilicity decrease upon addition of hydrogen molecules to $5^{12}6^8$ cavity in most cases (Figure 5). In both the cases, a little bit deviation is observed with the addition of four and five hydrogen molecules. Addition of six hydrogen molecules to $5^{12}6^8$ cage shows a more favorable trend for electrophilicity. We have calculated all global reactivity descriptors at MP2/6-31+G(d) level of theory and a similar trend was observed in both the levels (Table 6). Calculations at the MP2/6-31+G(d) give a better correlation compared to the B3LYP/6-31G(d) level. Electronegativity and electrophilicity decrease upon addition of hydrogen molecules onto the $5^{12}6^8$ cavity. The electronegativity (χ, eV), hardness (η, eV) and electrophilicity (ω, eV) values are shown in Table 6.

**Conclusion**

Computational studies reveal that cage structures $5^{12}$ and $5^{12}6^2$ may accommodate up to two hydrogen molecules and $5^{12}6^8$ may store up to six hydrogen molecules. For $5^{12}$ cage, encapsulation of first hydrogen molecule is a favorable process wherein $H_2$ strongly binds to the $5^{12}$ cage, whereas for $5^{12}6^2$, encapsulation of one or two hydrogen molecules is also energetically favorable process. The rest of the processes are not favorable. The MP2/6-31+G(d) single point energies predict better interaction energies than that calculated at the B3LYP/6-31G(d) level of theory.

Interestingly for $5^{12}6^8$ cage structure, the encapsulation of up to all six hydrogen molecules is a favorable process. The stability of the $H_2$-encapsulated cluster complexes has been understood both from the viewpoint of increasing hardness (η) trend and a subsequent decrease in the electrophilicity (ω). The interaction energy and the DFT based reactivity descriptors indicate that an increasing number of hydrogen molecules in the cage results in the enhanced stability of these complexes in most cases.

**Acknowledgements**

We thank Indo- EU (HYPOMAP) project and UGC for financial assistance. Authors would like to thank Dr. Hemanth Srivastava, Mr. Santanab Giri and Prof. Altaf Pandith for helpful discussions. One of us (PKC) would like to thank the DST, New Delhi for the J.C. Bose Fellowship.

**Supporting Information Available**

Optimized cartesian coordinates and total energies (hartrees) and NIMAG values calculated at B3LYP/6-31G (d) level of theory are given.